\documentclass[12pt,a4paper]{article}
\usepackage[utf8]{inputenc}
\usepackage[pdftex]{graphicx}
\usepackage{amssymb}
\usepackage{amsmath}
\usepackage{color}
\usepackage{cite}

\renewcommand{\vec}[1]{{\boldsymbol{#1}}}

\begin{document}

\title%
{%
On one effect of coronal mass ejections influence on the envelopes of hot Jupiters
}%
\author%
{%
 Zhilkin A.G.\thanks{E-mail: zhilkin@inasan.ru}, %
 Bisikalo D.V., %
 Kaigorodov P.V.\\[0.5cm] %
\textit{\small Institute of astronomy of the RAS, Moscow, Russia}\\%
}%
\date{}

\maketitle%

\begin{abstract}
It is now established that the hot Jupiters have extensive gaseous (ionospheric) envelopes, which expanding far beyond the Roche lobe. The envelopes are weakly bound to the planet and affected by strong influence of stellar wind fluctuations. Also, the hot Jupiters are located close to the parent star and therefore the magnetic field of stellar wind is an important factor, determining the structure their magnetosphere. At the same time, for a typical hot Jupiter, the velocity of stellar wind plasma, flowing around the atmosphere, is close to the Alfv\'en velocity. This should result in stellar wind parameters fluctuations (density, velocity, magnetic field), that can affect the conditions of formation of bow shock waves around a hot Jupiter, i. e. to switch flow from sub-Alfv\'en to super-Alfv\'en mode and back. In this paper, based on the results of three-dimensional numerical MHD modeling, it is confirmed that in the envelope of hot Jupiter, which is in Alfv\'en point vicinity of the stellar wind, both disappearance and appearance of the bow shock wave occures under the action of coronal mass ejection. The paper also shows that this effect can influence the observational manifestations of hot Jupiter, including luminosity in energetic part of the spectrum.
\end{abstract}

\section{Introduction}

One of the most important tasks of modern astrophysics is to study the mechanisms of mass loss by hot Jupiters. Hot Jupiters are exoplanets with a mass of the Jupiter mass order, whose orbits are located in close proximity to the parent star \cite{Murray2009}. The first hot Jupiter was discovered in 1995 \cite{Mayor1995}. The immediate vicinity to the parent star and the relatively large size of the hot Jupiters can lead to the expansion of atmosphere beyond the Roche lobe and the formation of an extended gaseous envelope. This process is accompanied by the formation of outflows from the Lagrange points vicinities  $L_1$ and $L_2$ \cite{Lai2010, Li2010}. The existence of such outflows is indirectly indicated by the excess absorption in the near ultraviolet range, observed among some planets \cite{Vidal2003, Vidal2008, BenJaffel2007, Vidal2004, BenJaffel2010, Linsky2010}. These conclusions are also supported by theoretical calculations in the frameworks of one-dimensional aeronomic models \cite{Murray2009, Yelle2004, Garcia2007, Koskinen2013, Ionov2017}.

The structure of the gaseous envelopes of hot Jupiters was studied by three-dimensional numerical modeling in the series of papers (see, e. g., \cite{Bisikalo2013hotjup, Cherenkov2014, Bisikalo2016, Cherenkov2017, Cherenkov2018, Bisikalo2018, Shaikhislamov2018, Cherenkov2019}). It was shown that depending on the parameters of model, hot Jupiters can form three main types of gaseous envelope. In the case of closed envelopes, the planet atmosphere is completely located inside its Roche lobe. In the case, where dynamic pressure of the stellar wind stops the outflow from vicinity of point $L_1$ outside the Roche lobe, quasi-closed envelopes are formed. Open envelopes are formed by outflows from Lagrange points $L_1$ and $L_2$ in the case when dynamic pressure of the stellar wind is not enough to stop them. The magnitude of mass loss rate significantly depends on the type of gaseous envelope formed.

In \cite{Arakcheev2017, Bisikalo2017} the results of three-dimensional numerical simulation of the flow structure in vicinity of hot Jupiter WASP 12b were presented, taking into account the influence of planet proper magnetic field. Calculations have shown that the presence of planet magnetic field can lead to an additional weakening of mass loss rate, compared to the pure gas-dynamic case. The analysis, carried out in \cite{Zhilkin2019}, showed that the magnetic field of the stellar wind is a very important factor, since many hot Jupiters are located in the sub-Alfv\'en zone of the stellar wind, where the magnetic pressure exceeds the dynamic one. In this case, the flow around process can be shockless \cite{Ip2004}. The paper \cite{Zhilkin2019} also shows that the absolute majority of hot Jupiters are near the boundary, which separates systems with sub-Alfv\'en and super-Alfv\'en flow regimes.

Various perturbations of the stellar wind can lead to significant changes in the structure of hot Jupiters gaseuos envelope and, consequently, to variations in mass loss rate. The most significant perturbation of the wind arise due to the huge ejections of a matter from a coronae of star --- coronal mass ejections (CMEs). In the papers \cite{Bisikalo2016, Cherenkov2017, Kaigorodov2019} using three-dimensional numerical simulations have shown that even in the case of a typical solar CME, the outer parts of hot Jupiter's asymmetric gaseous envelope outside its Roche lobe can be ripped off and carried into the interplanetary medium. This leads to a sharp increase in mass loss rate of hot Jupiter at the time of passing CME through it. 

In this paper, some interesting features of the CME interaction with magnetosphere of hot Jupiter, due to influence of the stellar wind magnetic field, are considered. The main attention is paid to the processes of formation and disappearance of bow shock wave in the envelope of hot Jupiter, located close to the Alfv\'en point of the stellar wind, due to changes in magnetic field of the stellar wind during the CME. The structure of this paper is organized as follows. Section 2 describes a simple CME model that takes into account the magnetic field. Section 3 analyzes the possible effects associated with the impact of CME on the magnetosphere of hot Jupiter. Section 4 discusses the main conclusions.

\section{MHD model of CME}

\begin{table}[t]
\begin{center}
\begin{tabular}{ccccc}
\hline\noalign{\smallskip}
Phase & 1 & 2 & 3 & 4 \\
\hline\noalign{\smallskip}
Duration (hours) & -- & 8.5 & 13 & 22 \\
$n/n_w$ & $1$ & $4$ & $0.6$ & $10$ \\
$T/T_w$ & $1$ & $5.07$ & $0.79$ & $0.30$ \\
$v/v_w$ & $1$ & $1.33$ & $1.44$ & $1.11$ \\
$B/B_w$ & $1$ & $2.25$ & $1.75$ & $1.13$ \\
$\lambda/\lambda_w$ & $1$ & $1.18$ & $0.63$ & $3.11$ \\
\noalign{\smallskip}\hline
\end{tabular}
\caption{Parameters (density, temperature, velocity, magnetic field and Alfv\'en Mach number) of stellar wind during the CME passage.}
\label{tb1}
\end{center}
\end{table}

Our numerical model of the stellar wind during the passage of CME in vicinity of a planet based on the measurements of the solar wind parameters at the Earth orbit, obtained by spacecraft ACE, WIND, SOHO, in May 1998 during the relevant event \cite{Farrell2012}. The averaged values of these parameters are given in the table~\ref{tb1}. 
The process of CME passing can be divided into four separate phases. The first phase corresponds to the state of unperturbed solar wind. The second phase begins with the passage of MHD front of the shock wave and is characterized by increase in the number density $n$ with respect to the unperturbed value $n_w$ by about 4 times. At the same time, speed increases by 1.3 times. Behind the shock wave front, the magnitude of magnetic field induction increases by 2.25 times. The shock wave is followed by a sheath of denser matter. The duration of this phase is approximately 8.5 hours. The third phase (early CME) starts with the passage of tangential MHD discontinuity, which propagates after the shock wave. Thus density falls about 2 times compared to the unperturbed value. Duration of this phase is 13 hours. Finally, the fourth phase (late CME) is characterized by a sharp increase in density (about 10 times with respect to unperturbed wind). However, this phase does not have a clearly defined boundary and, apparently, its beginning is not associated with the passage of any discontinuity. Duration of this phase is about 22 hours. After that, the wind parameters return to their original values.

\begin{figure}[t]
\centering
\includegraphics[width=0.95\textwidth]{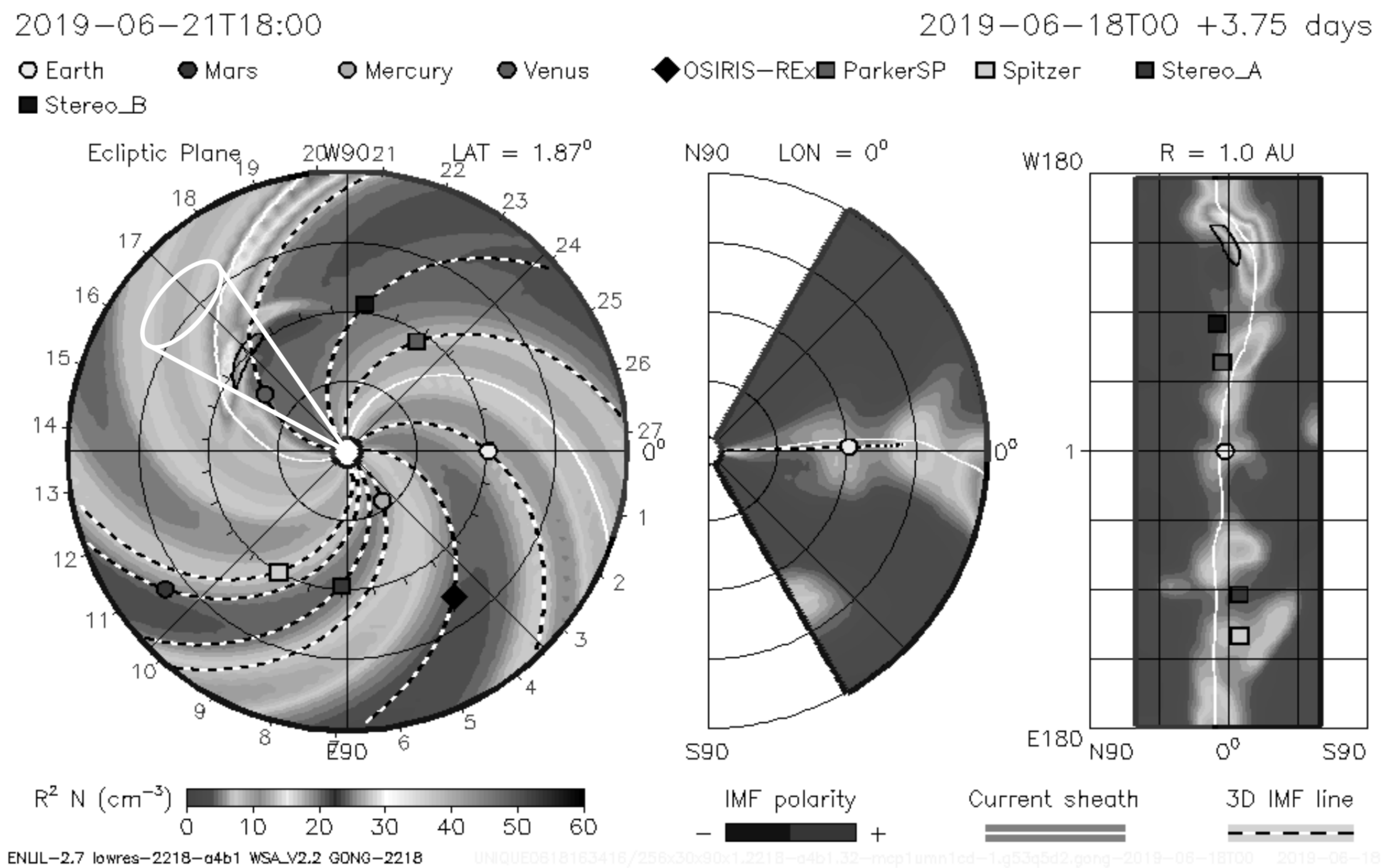}
\caption{Solar wind structure at the end of June 2019 according to spacecraft data. The positions of the inner planets, magnetic lines of force, and the configuration of the current sheet are shown. The CME cone, which passed in the vicinity of Venus, is shown. Taken from the site https://ccmc.Gsfc.nasa.gov/iswa.}
\label{fg-iswa}
\end{figure}

The real structure of the solar wind, obtained from spacecraft data at the end of June 2019, is shown in Fig.~\ref{fg-iswa}.
The current positions of all inner planets (including Mars), selected magnetic field lines, as well as the configuration of the current sheet, separating two sectors of the solar wind with different polarity of the heliospheric magnetic field are shown. At this time, the solar wind was observed passing CME, whose cone is indicated on the left panel of Fig.~\ref{fg-iswa}.
The ejection was directed towards Venus, which led to its interaction with that planet ionospheric envelope. The angle at ejection cone apex was approximately $27^\circ$.

\begin{figure}[t]
\centering
\includegraphics[width=0.75\textwidth]{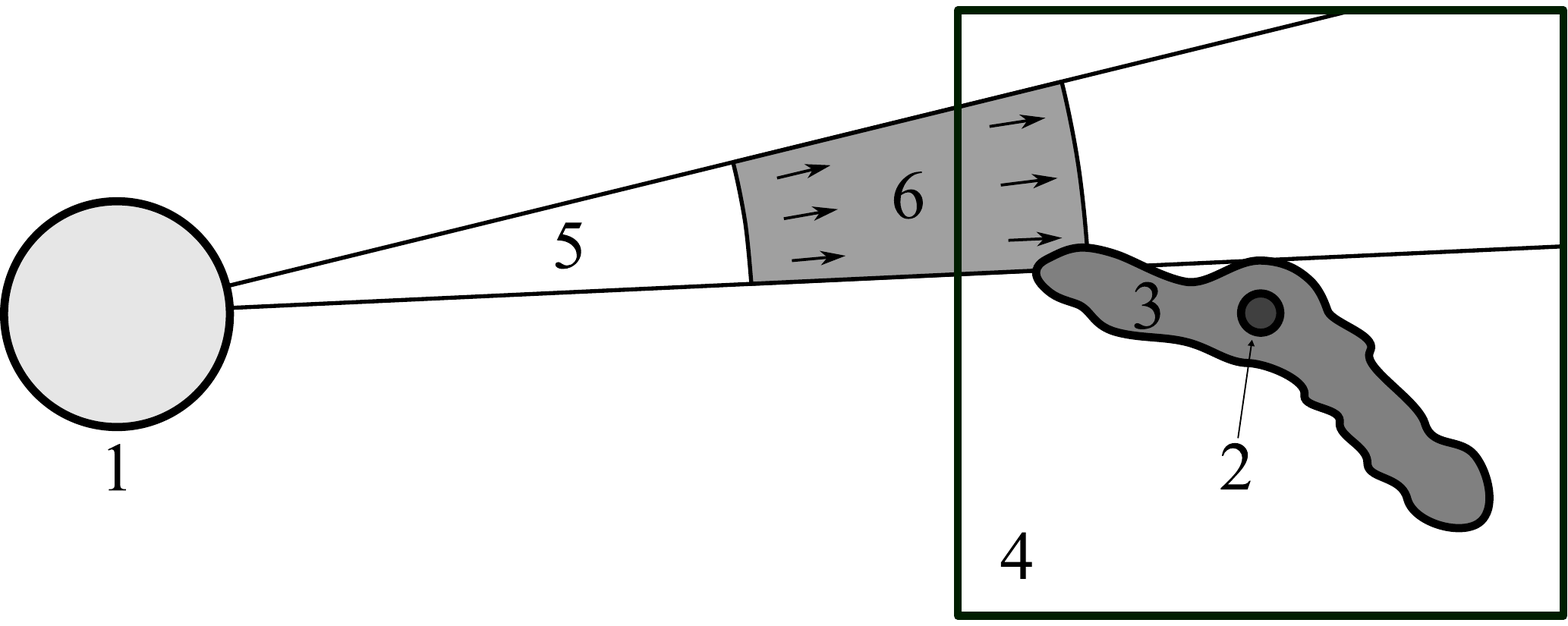}
\caption{Model of coronal mass ejection. The numbers are: 1 -- star, 2 -- planet, 3 -- gaseous envelope of the planet, 4 -- the calculation area, 5 -- ejection cone, 6 -- the area of spatial localization of the ejection.}
\label{fg1}
\end{figure}

Time-dependent factors $f_q(t)$ --- CME time profiles --- are used to describe the process of passing CME in vicinity of the planet. These factors determine the perturbations of stellar wind initial parameters at the observation point. In other words, for some parameter $q$, describing the state of stellar wind, we can write the following expression:
\begin{equation}\label{eq-1}
 q(\vec{r}, t) = f_q(t) q_w(\vec{r}).
\end{equation}
Here $q_w (\vec{r})$ describes the unperturbed steady state of the wind at a given point $\vec{r}$, and $q(\vec{r}, t)$ corresponds to the state of wind perturbed by the passage of CME. The functions $f_q(t)$, describing the time profile of CME in vicinity of the planet, for all values (density $n$, temperature $T$, velocity $v$ and magnetic field $B$) correspond to the parameters given in the table \ref{tb1}. 
Let us assume that the matter of which it is composed, moves from the star inside some cone \cite{Kaigorodov2019}. The corresponding flow pattern is schematically shown in Fig. \ref{fg1}.
The angle at the cone apex $\alpha$, as well as its orientation in space, which is determined by the angles $\theta$ and $\phi$ in the spherical coordinate system, are the parameters of our model. The interaction process of gaseous envelope of the hot Jupiter type planet with the matter of parent star stellar wind, taking into account the CME, can be described on the basis of relation \eqref{eq-1}. To do this, it's necessary at first to determine the parts of calculated area boundary, that intersect with the ejection cone. At these sites, the relations \eqref{eq-1} should be used to specify unsteady boundary conditions. To determine the ejection phase at different points in space, located at various distances from the star, we should have an information about the CME speed, which is not available in the experimental data. Therefore, as this velocity in numerical calculations, it is possible to approximate the gaseous velocity in this phase of CME. We emphasize that the perturbations of stellar wind parameters include purely hydrodynamic quantities (density, velocity, temperature) and magnetic field. Therefore, the relations \eqref{eq-1} remain valid in the case of MHD modeling.

To determine whether a given point of space, whose radius--vector is $\vec{r}$, falls into the region of a CME cone, we introduce a unit vector $\vec{a}$ along the axis of cone. Its components in the Cartesian coordinate system can be written as:
\begin{equation}\label{eq-2}
 a_x = \sin\theta \cos\phi, \quad
 a_y = \sin\theta \sin\phi, \quad
 a_z = \cos\theta.
\end{equation}
Then an angle $\beta$ between the axis of cone and the direction from apex of cone to the observation point is determined by follow relation:
\begin{equation}\label{eq-3}
 \cos\beta = \vec{a} \cdot \vec{n},
\end{equation}
where vector $\vec{n} = \vec{r} / r$.The observation point is inside the CME cone, if the follow condition is satisfied:
\begin{equation}\label{eq-4}
 \cos\beta > \cos(\alpha/2).
\end{equation}
In a non-inertial reference frame, associated with an orbiting planet, in these relations it is necessary to take into account that the azimuthal angle $\phi$ of ejection orientation changes in time according to this law:
\begin{equation}\label{eq-5}
 \phi = \phi_0 - \Omega (t - t_0),
\end{equation}
where $\Omega$ is the angular velocity of the planet's orbital rotation, $t_0$ -- some initial moment of time, and $\phi_0$ is the corresponding initial phase. As $t_0$ it is possible to choose, e. g., the moment of ejection entrance into the calculation area.

It should be noted that time profiles $f_q (t)$ can have a more general form \cite{Cherenkov2017}. Even for the Sun, CME intensities can vary significantly. For the stars of hot Jupiters, these variations may be even more pronounced. Within the framework of described model, with a more general approach, in functions $f_q(t)$ can vary both the relative changes of the parameters in phases and the duration of phases themselves. In addition, of piecewise constant functions, it is possible to use, e. g., piecewise linear functions instead. In this case, we can describe CMEs of various types, corresponding to slow, medium \cite{Mostl2014} and fast \cite{Liu2014} CMEs. 

Note also that the CME parameters represented in Table \ref{tb1},
are registered near the Earth orbit, where the shock wave, determining the beginning of the first phase, is almost purely gas-dynamic. This is due to the fact that magnetic field of the solar wind in this region is weak. In the sub-Alfv\'en zone of the solar wind, the shock wave at front boundary of CME is a fast MHD shock wave. Therefore, its parameters (in particular, the velocity of propagation of its front) may differ markedly from the parameters of corresponding purely gas-dynamic shock wave. 

\section{The features of CME passing in close proximity of hot Jupiters}

The analysis, carried out in \cite{Zhilkin2019}, allows to make a conclusion that in vicinity of almost all hot Jupiters, known to date, the speed of stellar wind is close to the Alfv\'en velocity. At the same time, many of them even can be in the sub-Alfv\'en zone, in which the stellar wind magnetic pressure exceeds its dynamic one. This means that the magnetic field is an important factor in study of the stellar wind, which flow around ionospheric envelope of hot Jupiter. Given factor should take into account, when both theoretical models constructing and interpretation of observational data are making.

\begin{figure}[t]
\centering
\includegraphics[width=0.9\textwidth]{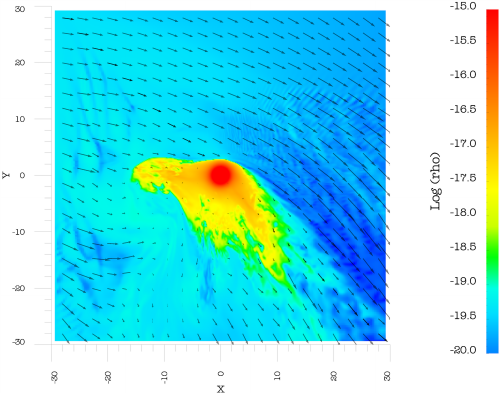}
\caption{Density and velocity distribution in the plane of a hot Jupiter orbit under the sub-Alfv\'en flow regime. The solution is presented at time $0.27 P_\text{orb}$ from the beginning of modelling.}
\label{strong_rho}
\end{figure}

\begin{figure}[t]
\centering
\includegraphics[width=0.9\textwidth]{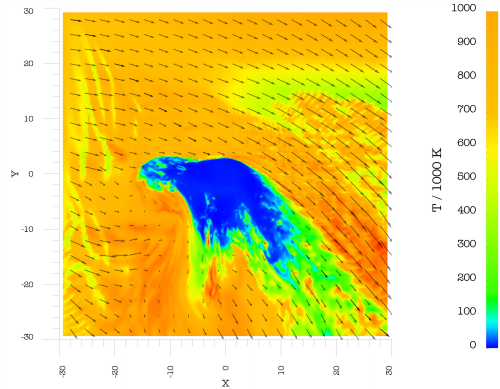}
\caption{Temperature and velocity distribution in the plane of a hot Jupiter orbit under the sub-Alfv\'en flow regime. The solution is presented at time $0.27 P_\text{orb}$ from the beginning of modelling.}
\label{strong_t}
\end{figure}

\begin{figure}[t]
\centering
\includegraphics[width=0.9\textwidth]{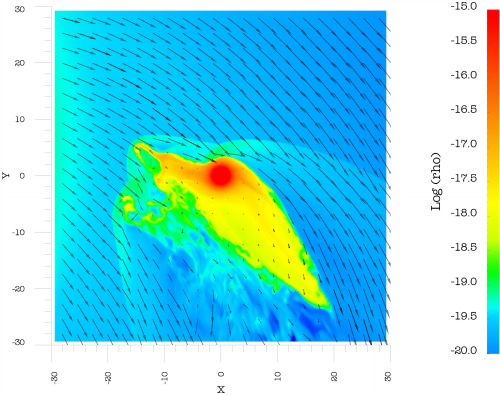}
\caption{Density and velocity distribution in the plane of a hot Jupiter orbit under the super-Alfv\'en flow regime. The solution is presented at time $0.26 P_\text{orb}$ from the beginning of modelling.}
\label{weak_rho}
\end{figure}

\begin{figure}[t]
\centering
\includegraphics[width=0.9\textwidth]{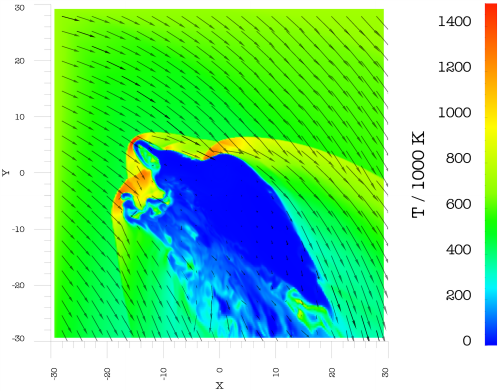}
\caption{Temperature and velocity distribution in the plane of a hot Jupiter orbit under the super-Alfv\'en flow regime. The solution is presented at time $0.26 P_\text{orb}$ from the beginning of modelling.}
\label{weak_t}
\end{figure}

Since the Alfv\'en velocity does not exceed the fast magnetosonic velocity, for hot Jupiters, whose orbits are located in the sub-Alfv\'en zone, wind speed is less than the fast magnetosonic velocity. In pure gas dynamics, such a case corresponds to subsonic flow around the body, in which the bow shock wave does not formed. A similar situation is realized in magnetic hydrodynamics. This means that the flow of stellar wind around the hot Jupiter, located in the sub-Alfv\'en zone, must be shockless
\cite{Ip2004}. 
On Fig. \ref{strong_rho}--\ref{weak_t} the results of three-dimensional numerical simulation of the flow structure in vicinity of the hot Jupiter HD 209458b \cite{Charbonneau2000} are presented. The numerical model described in \cite{Zhilkin2019} was used for calculations. The main parameters of the model corresponded to the values, used in our previous papers (see, e. g., \cite{Bisikalo2013hotjup}). The star HD 209458 of G0 spectral class has a mass of $M_s = 1.15 M_\odot$ and a radius of $R_s = 1.2 R_\odot$. The proper rotation of the star is characterized by a period of $P_\text{rot} = 14.4$ days, that corresponds to the angular velocity $\Omega_s = 5.05 \cdot 10^{-6}~\text{s}^{-1}$ or the linear velocity at the equator $v_\text{rot} = 4.2~\text{km}/\text{s}$. The planet has a mass of $M_p = 0.71 M_\text{jup}$ and a photometric radius of $R_p = 1.38 R_\text{jup}$, where $M_\text{jup}$ and $R_\text{jup}$ -- the mass and radius of Jupiter. The major semi-axis of the planet orbit $A = 10.2 R_\odot$, which corresponds to orbital period around the star $P_\text{orb} = 84.6$ hours.

At the initial moment of time around the planet was given a spherically symmetric 
isothermal atmosphere, in which the density distribution was determined from the condition of hydrostatic equilibrium. The radius of atmosphere was determined from the condition equilibrium in pressure with the matter of stellar wind. Temperature of atmosphere set to $T_\text{atm} = 7500~\text{K}$, and the number density at photometric radius $n_\text{atm} = 10^{11}~\text{cm}^{-3}$. 

As parameters of the stellar wind were used the corresponding values 
for the solar wind at a distance of $10.2 R_\odot$ from center of the Sun 
\cite{Withbroe1988}: temperature $T_w = 7.3\cdot 10^5~\text{K}$, speed $v_w 
= 100~\text{km}/\text{s}$, number density $n_w = 10^4~\text{cm}^{-3}$. The wind magnetic field was given by the formulas, mentioned in \cite{Zhilkin2019}. 
The average magnetic field on the surface of star was set to $B_s = 
0.5~\text{Gs}$ (strong field, Fig.~ \ref{strong_rho} and~\ref{strong_t})
and $B_s = 0.01~\text{Gs}$ (weak field, Fig.~\ref{weak_rho} and~\ref{weak_t}).
Given that the radius of star is slightly larger than the Sun radius, magnitude of 0.5~Gs almost corresponds to the average magnetic field at surface of the Sun --- 1~Gs, when comparing the respective magnetic moments of stars.

In our numerical model, we also took into account the planet proper magnetic field. By thus, we assumed that the value of magnetic moment of hot Jupiter 
HD 209458b $\mu = 0.1\, \mu_\text{jup}$, where 
$ \mu_\text{jup} = 1.53 \cdot 10^{30}~\text{Gs} \cdot \text{cm}^3$ -- magnetic moment of Jupiter. This value is consistent with observational 
\cite{Kislyakova2014}, and with theoretical \cite{Stevenson1983} estimates.
In the model it was assumed that the proper rotation of planet is synchronized with the orbital one, and the axis of proper rotation is collinear with the orbital axis.

Calculations were carried out until $t=0.27 P_{\text{orb}}$ in the case of a strong magnetic field of wind (Fig.~\ref{strong_rho},~\ref{strong_t}) and $t=0.26 P_{\text{orb}}$ in the case of a weak wind magnetic field (Fig.~\ref{weak_rho}, \ref{weak_t}).
As seen in Fig.~ \ref{strong_rho} and~\ref{strong_t}, the process of stellar wind interaction with ionospheric envelope of the planet in case of a strong magnetic field is shock-less. The bow shock wave forms neither around the planet atmosphere, nor around the matter, ejected from the point $L_1$. This is clearly seen in both the density and temperature distribution. The wind magnetic field is so strong that it prevents a free movement of plasma in transverse direction to the field lines. Therefore, the ejected matter moves towards the star mainly along magnetic lines of wind field. So we can say that in this process, the electromagnetic force due to the magnetic field of the wind plays an important role, compared to the star gravity, centrifugal force and Coriolis force.

In contrast, in the second model (weak magnetic field) as a result of the interaction of stellar wind with ionospheric envelope of the planet, a bow shock wave forms. This is clearly seen in Fig. \ref{weak_rho} and \ref{weak_t}.
At the same time, it can be argued that it consists of several intersecting shock waves, one of which occurs when the wind interacts with the jet matter from the inner Lagrange point $L_1$, and the rest --- directly with the planet atmosphere and the plume of matter behind it. Inside the Roche lobe of the planet, magnetic field preserves a dipole structure. Since in this case, magnetic field of stellar wind is weak and does not play any significant dynamic role, the flow structure in the envelope is close to the purely gas-dynamic case.

Thus, on the basis of numerical calculations, which results are represented in Fig. \ref{strong_rho}--\ref{weak_t},
it can be concluded that the decrease in magnetic field of wind really leads to formation of a bow shock wave. Since hot Jupiters are located near the Alfv\'en point of stellar wind, this, in particular, means that even relatively small fluctuations in the flow around stream can lead to disappearance or, conversely, to appearance of shock waves around the planet.

Let us now consider the MHD features of CME interaction process with ionosphere envelope of a hot Jupiter. 
In Table \ref{tb1}, the last line shows changes in value of the Alfv\'en Mach number 
\begin{equation}\label{eq-6}
\frac{\lambda}{\lambda_w} = \sqrt{\frac{n}{n_w}} \frac{v}{v_w} \frac{B_w}{B}
\end{equation}
at different phases of CME passage. As we can see from the table, the $\lambda$ value changes in a non-monotonic way. At the first phase, $\lambda$ slightly exceeds the unperturbed value of $\lambda_w$, at the second phase $\lambda$ becomes smaller than unperturbed value $\lambda_w$, at the third phase $\lambda$ again 
increases sharply and more than three times the unperturbed value $\lambda_w$.

If a planet is deep in the sub-Alfv\'en zone or, conversely, far in the 
super-Alfv\'en zone, the nature of flow during the passage of CME is not 
changes. With a strong wind field, it will be shock-less, as in the results shown in Fig.~\ref{strong_rho} and~\ref{strong_t}.
In the case, where the wind field is weak, the whole process from beginning to the end will be accompanied by formation of bow shock waves, as in calculation shown in Fig.~\ref{weak_rho} and~\ref{weak_t}) and how
this was observed in purely gas-dynamic calculations \cite{Bisikalo2016, 
Cherenkov2017, Kaigorodov2019}. However, if the planet orbit is near 
the Alfv\'en point, CME interaction process with magnetosphere can be more complex and interesting. Recall that for hot Jupiters it is there must be a very common case of \cite{Zhilkin2019}.

Let's assume that such a planet is near the Alfv\'en point, but from the outside 
sub-Alfv\'en wind zone. Then, at the second phase, a flow regime should 
to remain shock-less, because in this phase the Alfv\'en Mach number is smaller than
the unperturbed value of $\lambda_w$. At the first and third phases, the Alfv\'en Mach number, on the contrary, increases compared to the unperturbed value. Depend on specific situation this may be enough to make the flow rate become greater than the fast magnetosonic speed, either at the third phase CME, or immediately at the first and the third phases. In the first case, the third phase of CME will have a bow shock wave, which disappear again in the end of entire process and returns system to the original unperturbed condition. In the second case, the shock wave occurs already in the first phase, at the second phase it disappears, then reappears at the third phase, and finally, disappears after CME passage. 

Let us now suppose that hot Jupiter is near the Alfv\'en point, but with 
the sides of the super-Alfv\'en wind zone. Then, flow mode can be switched to 
the second phase of CME passage, when the Alfv\'en Mach number decreases by 
compared to the unperturbed value. That might be enough to change the flow mode to shock-less one, when the flow rate becomes less than the fast magnetosonic speed and a bow shock wave has not formed, as in our calculations above (see Fig.~ \ref{strong_rho} and \ref{strong_t}). 
As a result of changing the flow mode, the bow shock wave can "turn off" for a while, and then "turn on" again after the second ejection phase is over.

\section{Conclusion}

The paper deals with the influence of perturbation of stellar wind parameters caused by passage of a coronal mass ejection, on the nature of flow near hot Jupiter. It has been shown that when hot Jupiter orbits is close to the Alfv\'en point, passing through CME can cause a temporary appearance or disappearance of a shock wave, as the flow can switch from sub-Alfv\'en to super-Alfv\'en mode and back.
This is important, because many hot Jupiters may be in the sub-Alfv\'en zone of a stellar wind or around the boundaries of this zone \cite{Zhilkin2019}.

As has been shown in~\cite{Bisikalo2016, Cherenkov2017, Kaigorodov2019}, the passage through CME can significantly change the structure of envelope and lead to an increased of mass loss rate by hot Jupiter, which should affect its observational manifestations.
However, the occurrence or disappearance of a shock wave can also lead to the observable effects. One of the possible manifestations, associated with the presence of a shock wave, may be radiation in the X-ray part of the spectrum. As seen in Fig.~ \ref{weak_t},
the temperature in shock waves in front of the planet envelope can reach quite high values (up to $1.5 \cdot 10^6$~K), while the average thermal velocity of particles in the gas (the speed of sound) is $\sim 144$ km/s.
The velocity of particle collision on the shock wave front also depends on velocity jump on it, which is $\sim 160$ km/s in our solution.
This gives an average particle collision velocity of $\sim 300$ km/s when passing the shock wave front.
The collision of protons, which has such velocities, should lead to a relatively hard X-ray radiation with an energy of $\sim 1$ keV, associated with the presence of a shock wave. 

Given the relatively high luminosity of a bow shock wave, effect of its occurrence/disappearance can be detected during X-ray observations of exoplanets at moment of CME passing. An example of such observations presented in ~\cite{Czesla2019}. Results of observations in the X-ray spectrum of a flare in the CVSO~30 system, which includes a hot Jupiter with a mass of $\sim 3.6 M_{jup}$ and an orbital period of $\sim 0.44$ days, were carried out in two ranges --- soft ($0.1\div 1$~keV) and hard ($1\div 9$~keV). In the hard range, a short-term fall in luminosity was recorded $\sim 2.7$ hours after the flare begining. Unfortunately, the characteristics of wind and magnetic field of the star CVSO~30 are unknown, but it can be assumed that the observed darkening in the X-ray range may be due to the transition of flow around the hot Jupiter from the super-Alfv\'en to the sub--Alfv\'en regime and the disappearance of a bow shock wave. If this is indeed the case, then we have the opportunity to use this data as a means of diagnosing the stellar wind. In fact, if, for example, we divide the distance between planet and the surface of star by time before dip in a light curve, we get an average speed of $\sim 60$ km/s, which is quite consistent with the typical wind speed at this distance for the Solar-type star. Additionally, we note that the paper~\cite{Czesla2019} indicates that at the same time, according to AAVSO, the darkening of this star in the optical range was recorded.

Another possible observational manifestation of flow transition from shock to shock-less mode and back may be caused by change in charge exchange between plasma of stellar wind and gas of atmosphere of a hot Jupiter. This process leads to appearance of high-energy particles in gas and to the corresponding broadening of absorption lines in atmosphere of hot Jupiter~\cite{Kislyakova2014}.
The disappearance of a shock wave should lead to density fall in stellar wind matter, directly interacting with the atmosphere of hot Jupiter, and to a corresponding decrease in charge exchange. A change in absorption lines during CME passage can give an additional information about both the properties of atmosphere of hot Jupiter and the parameters of stellar wind.

Coronal mass ejections occur quite often, especially in young stars, they are an important factor affecting the long-term evolution of hot Jupiters.
As shown above, passing through the CME can lead to a short-term decrease or increase in the X-ray flux, associated with a change in the flow regime near hot Jupiter.
Potentially, this effect can be observed not only for transit hot Jupiters, which gives a unique opportunity for discovery of exoplanets, whose detection is impossible in other ways. In addition, an analysis of changes in the X-ray flux makes it possible to estimate the parameters of stellar wind of distant stars, which is also very difficult to do in other ways.

The work was prepared with the support of The Russian Foundation for basic research 
(RFBR grant No. 18-02-00178). The work was carried out using the equipment of the center for collective use "Complex of modeling and data processing of research facilities mega-class" SIC "Kurchatov Institute", http://ckp.nrcki.ru/, as well as computing tools MSC RAS.

\end{document}